\documentclass{emulateapj}

\newcommand{\hMpc}{{\ifmmode{h^{-1}{\rm Mpc}}\else{$h^{-1}$Mpc}\fi}}
\newcommand{\hkpc}{{\ifmmode{h^{-1}{\rm kpc}}\else{$h^{-1}$kpc}\fi}}
\newcommand{\hMsun}{{\ifmmode{h^{-1}{\rm {M_{\odot}}}}\else{$h^{-1}{\rm{M_{\odot}}}$}\fi}}
\newcommand{\Msun}{{\ifmmode{{\rm {M_{\odot}}}}\else{${\rm{M_{\odot}}}$}\fi}}

\shortauthors{Gottl\"ober and Yepes}

\shorttitle{Shape, spin and baryon fraction of clusters}

\begin{document}

%%--------------------------------------------------------
\title{Shape, spin and baryon fraction of clusters in the MareNostrum Universe}
%%--------------------------------------------------------
\author{Stefan Gottl\"ober}
\affil{Astrophysikalisches Institut Potsdam,
An der Sternwarte 16, 14482 Potsdam, Germany}
\email{sgottloeber@aip.de}

\and

\author{ Gustavo Yepes} 
\affil{Grupo de Astrof\'\i sica,
Universidad Aut\'onoma de Madrid, Madrid E-28049, Spain}
\email{gustavo.yepes@uam.es}
%%
% \date{\today}
%%

%%
%\maketitle
%%

\begin{abstract}
  The {\em MareNostrum Universe} is one of the largest cosmological
  SPH simulation done so far. It consists of $1024^3$ dark and
  $1024^3$ gas particles in a box of 500 $h^{-1}$ Mpc on a side.  Here
  we study the shapes and spins of the dark matter and gas components
  of the 10,000 most massive objects extracted from the simulation as
  well as the gas fraction in those objects. We find that the shapes
  of objects tend to be prolate both in the dark matter and gas. There
  is a clear dependence of shape on halo mass, the more massive ones
  being less spherical than the less massive objects.  The gas
  distribution is nevertheless much more spherical than the dark
  matter, although the triaxiality parameters of gas and dark matter
  differ only by a few percent and it increases with cluster mass. The
  spin parameters of gas and dark matter can be well fitted by a
  lognormal distribution function.  On average, the spin of gas is 1.4
  larger than the spin of dark matter.  We find a similar behavior for
  the spins at higher redshifts, with a slightly decrease of the spin
  ratios to 1.16 at $z=1.$ The cosmic normalized baryon fraction in
  the entire cluster sample ranges from $Y_b = 0.94$, at $z=1$ to $Y_b
  = 0.92$ at $z=0$.  At both redshifts we find a slightly, but
  statistically significant decrease of $Y_b$ with cluster mass.
\end{abstract}
%%
%%--------------------
\section{Introduction}
%%--------------------
%%

According to the standard scenario cosmic structures form by the
gravitational collapse of density fluctuations.  This collapse is
mainly determined by the dark matter (DM) which contributes 85\% to
the total matter density in the universe. Baryonic matter follows the
DM and forms visible objects like galaxies and clusters inside the DM
halos.

Shape and angular momentum are two important characteristics of halos.
The shape of DM halos has been already widely studied, mainly by means
of N-body simulations (e.g. \cite {allgood:2006}, \cite {Maccio:2006},
\cite {bett:2006} and references therein).  \cite{avila:2005} discuss
the dependence of the shape of galaxy sized halos on environment.
\cite{bailin:2005} studied the internal shape of DM halos. Clusters of
galaxies are the most recently formed objects in the universe. Most of
the gas has not had time to cool. Since gas contributes only 15 \% of
the total mass one would expect that the gas follows DM and both
distribution should be similar. In order to quantify this similarity,
we have characterized the shape of both components in halos obtained
from a large non-radiative cosmological gasdynamical simulation.
Further we have compared the baryon fraction in clusters with the
cosmic baryon fraction.

The origin, evolution and distribution of DM halo spins have been
widely discussed in the past also (eg. \cite{vitvi:2002},
\cite{bullock:2001}, \cite {Maccio:2006}, \cite {bett:2006}). In
galactic size halos the angular momentum of the gas component is
important for the understanding of disk formation. Using gasdynamical
simulations \cite{vdbosch:2002}, \cite{chen:2003}, \cite{sharma:2005}
have studied the spin of the gas component in galactic halos and found
that  at redshift $z=0$  the gas component, on average, has a larger
spin than DM.  Here we will discuss briefly the relation between the
different definitions of the spin parameter and study the distribution
of spin parameters of DM and gas components in our  numerical clusters.

Most of the previous studies on shape and spin of halos used either
collisionless dark matter simulations with large number of particles
and relatively large volumes or gasdynamical (dissipative and non
dissipative) simulations in small volumes with a small number of
objects. The advantage of our study is that we have both gas and dark
matter components in the MareNostrum SPH simulation and a large
cluster sample (more than 10,000) to perform statistics, thanks to the
large computational volume and the high number of particles of this
simulation.

\begin{figure*}[t]
\plottwo{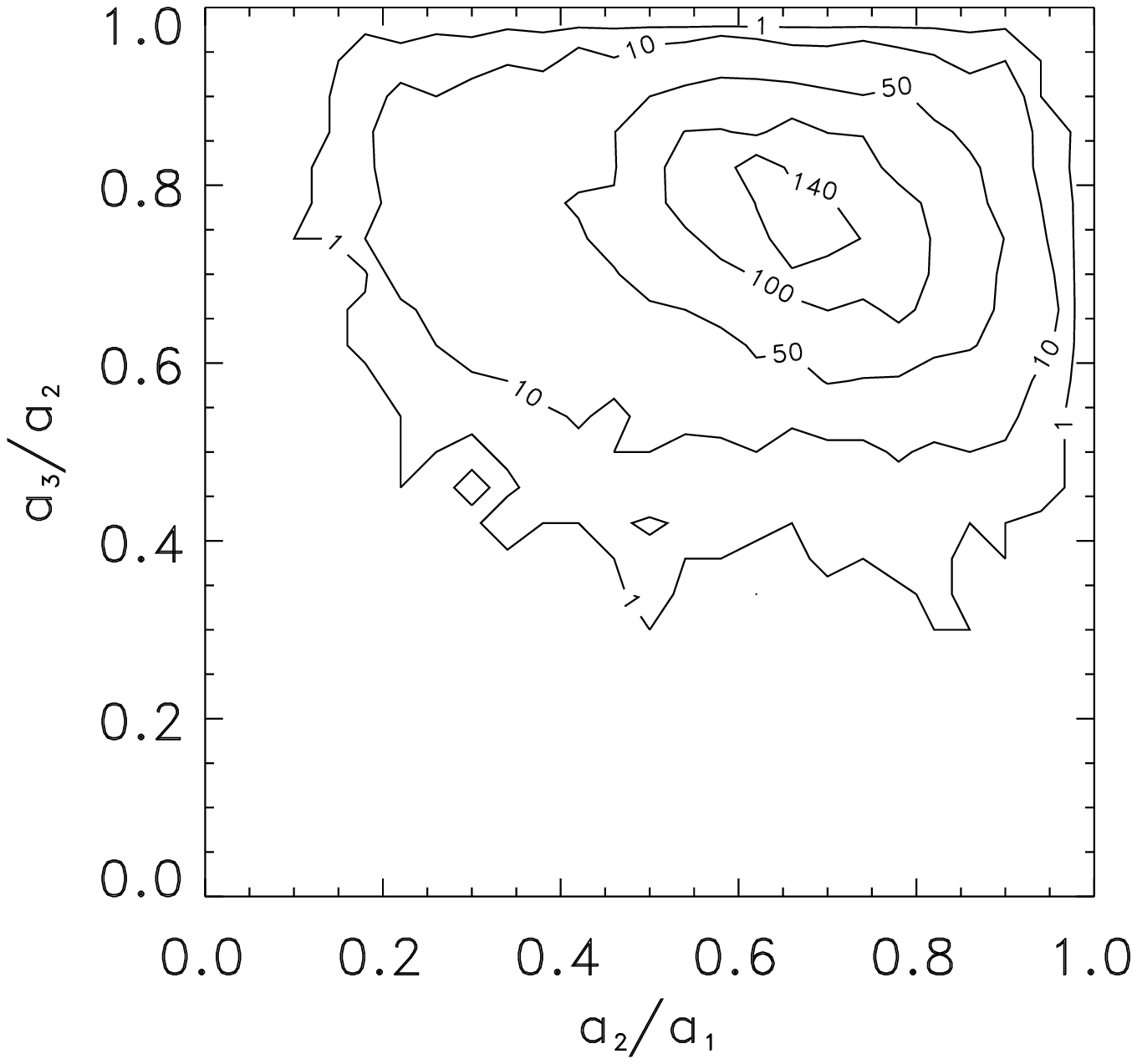}{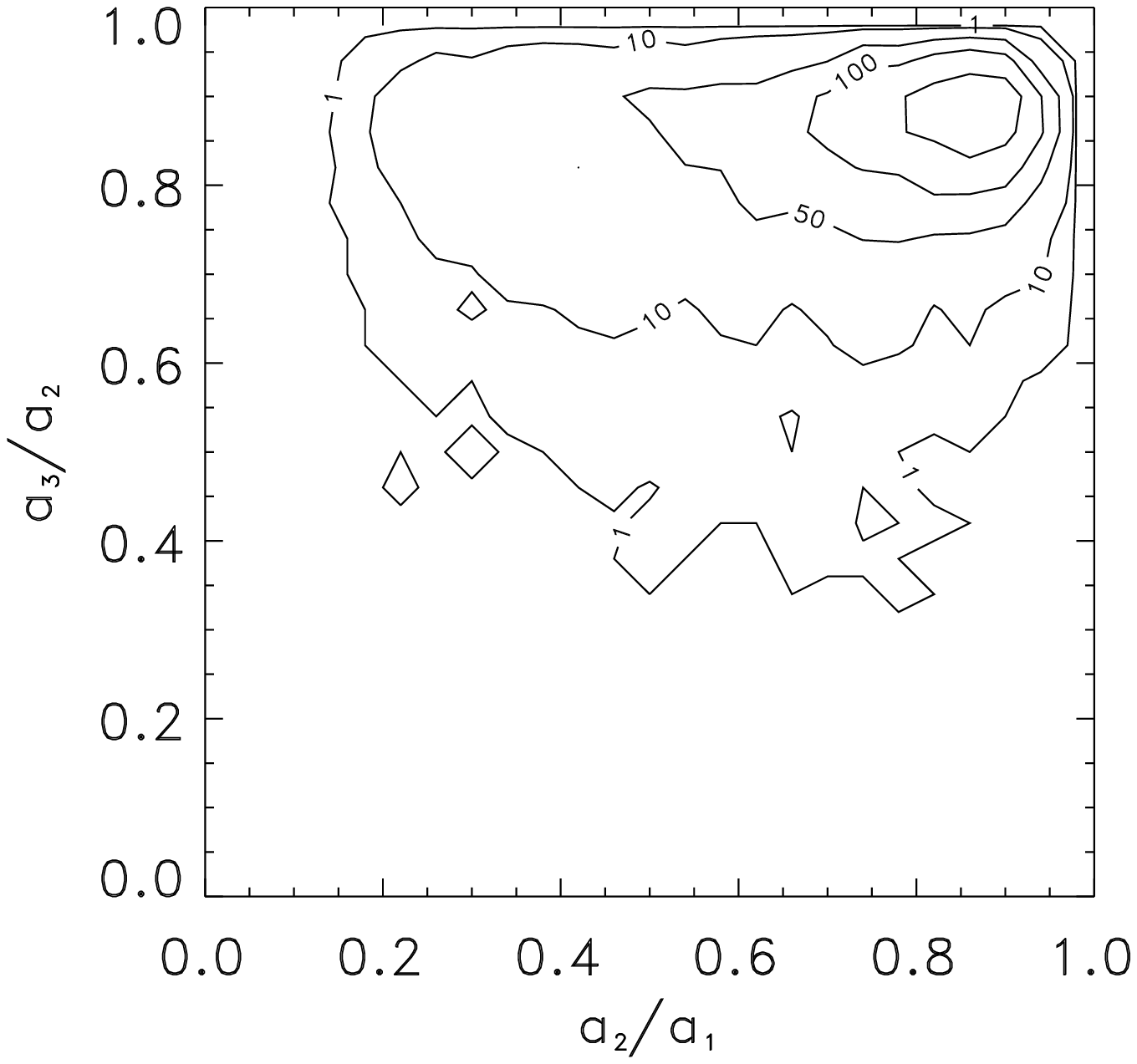}  
\caption{\label{plot:shape}
  {\it Left}: Shape of the dark matter distribution in clusters with
  total masses larger than $5 \times 10^{13} \hMsun$, $a_1$, $a_2$,
  $a_3$ are the three main axes  of the clusters.  {\it Right}: The
  same for the shape of the gas distribution. }
\end{figure*}

%%
%%---------------------------------------
\section{Numerical Simulation}
\label{sec:simu}
%%---------------------------------------
%%

This simulation, dubbed {\sc The MareNostrum Universe}, was performed
with the entropy conserving {\sc Gadget2} code \citep{Springel05} on
the MareNostrum supercomputer at the Barcelona Supercomputer Center
using the equivalent of about 29 years on a single CPU. It followed
the non linear evolution of structures in gas and dark matter (DM)
from $z=40$ to the present epoch ($z=0$) within a comoving cube of
$500\hMpc$ edges.  We assumed the spatially flat concordance
cosmological model with the following parameters: the total matter
density $\Omega_m=0.3$, the baryon density $\Omega_b=0.045$, the
cosmological constant $\Omega_\Lambda=0.7$, the Hubble parameter
$h=0.7$, the slope of the initial power spectrum $n=1$ and the
normalization $\sigma_8=0.9$. The power spectrum used to generate the
initial conditions for the simulation was kindly provided by Wayne Hu
in a tabulated form. It was obtained from a direct integration of the
Boltzmann code for the parameters described above. We did numerical
interpolation to compute the contribution of the different Fourier
modes.  Both components, the gas and the DM, were resolved by
$1024^3$ particles, which resulted in a mass of $8.3 \times
10^9\hMsun$ for the DM particles and $1.5 \times 10^9\hMsun$ for the
gas particles, respectively.

{\sc Gadget2} uses the TREEPM algorithm on a homogeneous Eulerian grid
to compute large scale forces by the Particle-Mesh algorithm.  In this
simulation we employed $1024^3$ mesh points to compute the density
field from particle positions and FFT to derive gravitational forces.
Within {\sc Gadget2} the equations of gas dynamics are solved by means
of the Smoothed Particle Hydrodynamics (SPH) method in its entropy
conservation scheme.  We did not include dissipative or radiative
processes or star formation.  The spatial force resolution was set to
an equivalent Plummer gravitational softening of $15 \;h^{-1}$
comoving kpc.  The SPH smoothing length was set to the distance to the
40$^{th}$ nearest neighbor of each SPH particle.

The {\sc MareNostrum universe} is part of a series of simulations we
have performed during the last years. To this end, we had generated a
realization of the $\Lambda CDM$ power spectrum with $2048^3$
particles and then decreased the mass resolution in the whole box to
$256^3$, $512^3$ and $1024^3$ particles respectively. The lower
resolution ($2\times 256^3$, $2\times512^3$) SPH simulations have been
used to study properties of galaxy clusters \citep{yepes:2004} and the
shape-alignment relation of clusters
\citep{faltenbacher:2005,basilakos:2006}. The MareNostrum Universe
have been recently used to analyzed the entropy profiles of the gas
and DM in galaxy clusters \citep{faltenbacher:2006}.  Due to
computational limitations, we could not yet simulate the evolution of
the full box with the maximal possible resolution of the initial
conditions, $2\times 2048^3$ particles. However, using the multi mass
technique described in \cite{klypin:2001} selected individual clusters
have already been re-simulated at this resolution
\citep{ascasibar:2006}.

It is a challenge to find within a distribution of 2 billion particles
all structures and substructures and to determine their properties.
Here we have used a newly developed parallel version of the
hierarchical friends-of-friends (FOF) algorithm \citep{klypin:1999}.
In a first step we construct the minimum spanning tree for the
distribution of gas and DM particles. After topological sorting we get
a cluster-ordered sequence from which we can easily extract FOF
clusters by simple cutting the tree at the desired linking length.  We
use a basic linking length of 0.17 of the mean interparticle
separation to extract the FOF clusters \citep{gottloeber:2006b}. We
divide this linking length by $2^n$ ($n=1,3$) to find substructures
and in particular the centers (density peaks) of our objects. We were
running the minimum spanning tree and the FOF analysis independently
over DM and gas particles to find their distribution.  Using a linking
length of 0.17 at redshift $z=0$ we have identified more than 2
million objects with more than 20 DM particles which closely follow a
Sheth-Tormen mass function \citep{gottloeber:2006a}.

To determine the shape and spin of the clusters we have selected a
subsample of more than 10,000 clusters and groups with masses larger
than $5 \times 10^{13}\hMsun$. The lower mass threshold corresponds to
clusters rsp. massive groups with about 5000 gas particles and 5000
dark matter particles.  As mentioned above the objects were identified
independently from the gas and DM distributions.  Due to the large
number of particles in this simulation we can unambiguously match DM
and gas halos with the same center of mass to one cluster.  In the
following we will study these two components of the clusters in more
detail.

%%
%%------------------------------
\section{Shapes}
\label{sec:shapes}
%%------------------------------
%%

Using the FOF method one extracts rather complex objects from the
simulation which are characterized by an iso-density surface given by
the linking length. In first approximation these objects can be
characterized by three-axial ellipsoids. The shape and orientation of
the ellipsoids can be directly calculated as eigenvectors of the
inertia tensor of the given object. Then the shape is characterized by
the ratios between the lengths of the axes $a_1 \geq a_2 \geq a_3$.

We ran the FOF algorithm independently over the DM and gas particles
to determine the shape of the distribution of the two components. In
Fig. \ref{plot:shape} we show the shape of the dark matter and gas
distribution in clusters with total masses larger than $5 \times
10^{13} \hMsun$. The ratios $a_3/a_2$ and $a_2/a_1$ of the clusters
have been sampled into 25 bins of size 0.04. The plot shows a clear
difference between the shapes of DM (left) and gas (right)
distribution.  DM halos are centered at ratios (0.7, 0.75) whereas for
the gas halos the center is at (0.85, 0.9), ({\it i.e.} the gas halos
are much more spherical).  Both shapes tend to be more prolate, i.e.
$a_3/a_2 > a_2/a_1$. This could be the result of merging along a
preferred direction, the large scale filaments
\citep{faltenbacher:2005}.
\begin{figure}
\plotone{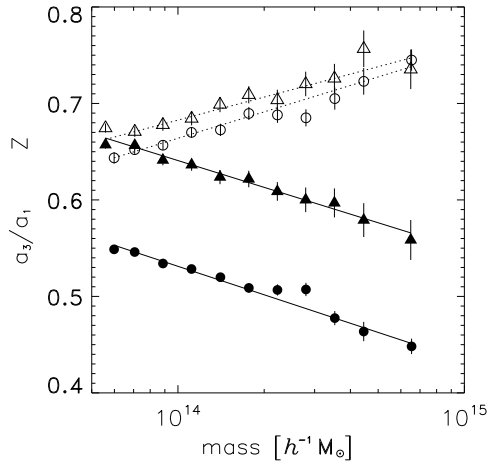}
\caption{\label{plot:triaxial}
Ratio of
  the minor to major axes $a_3/a_1$ (filled symbols) and triaxiality
  parameter $Z$ (open symbols) for the DM (circles) and gas (triangles)
  distribution in our numerical clusters}
\end{figure}

The exact position of the maximum in left and right panel of Fig.
\ref{plot:shape} depends on the low mass cut-off in the cluster
sample, but the qualitative behavior is independent of it. In
\ref{plot:triaxial}, we show how the ratio of the minor to major axis
$a_3/a_1$ of DM (filled circles) and gas (filled triangles) halos
depend on cluster mass. Note, that the dispersion in the values of the
ratios $a_3/a_1$ is large (within one bin the $1\sigma$ scatter is
0.14). This can be seen in Fig.  \ref{plot:shape} for the ratios
$a_2/a_1$ and $a_3/a_2$.  However, due to the large number of objects
per mass bin the standard deviation of the average within each bin
shown in Fig.  \ref{plot:triaxial} is small and sometimes smaller than
the symbol size.  The solid line is a linear fit in the
semi-logarithmic plot $a_3/a_1 = b + c \ln(M)$, where the mass $M$ is
given in $\hMsun$ and $b$ and $c$ are (1.904, -0.0426) and (1.932,
-0.0401) for DM and gas, respectively. A similar trend (but higher
$a_3/a_1$ have been found by \cite{kasun:2005} in the Hubble Volume
N-body simulation.

The cluster shape can be further characterized by the triaxiality
parameter $Z = (a_1-a_2)/(a_1-a_3)$ introduced by \cite{binney:1985}.
$Z=1$ corresponds to prolate ellipsoids of revolution (prolate
spheroids), while $Z=0$ corresponds to oblate ones. In Fig.
\ref{plot:triaxial} the triaxiality parameter $Z$ is shown for the DM
(open circles) and gas halos (open triangles). The dotted line is a
fit $Z = d + f \ln(M)$ with $(d,f) = ( -0.61, 0.039)$ for DM and
(-0.43, 0.034) for gas halos. Contrary to the ratio $a_3/a_1$ there is
only little difference between the triaxiality parameter $Z$ of the DM
and gas halos. The younger the halos are (i.e. more massive) the more
prolate they look.

%%
%%------------------------------
\section{Spin}
\label{sec:spin}
%%------------------------------
%%

In  N-body numerical simulations the identified clusters can be characterized
by their mass velocity and angular momentum or spin.  Originally the
spin parameter $\lambda$ was introduced by \cite{peebles:1971} as
the ``convenient dimensionless number''
\begin{equation}
\lambda =\frac{J |E|^{1/2}}{GM^{5/2}},
\label{lambda_peebles}
\end{equation}
where $J$ is the total angular momentum of the object, $E$ its total
energy and $M$ its mass. Following \cite{paddy:1993} one can
interprete $\lambda$ as the ratio of the angular velocity $\omega$ of
the system to the angular velocity $\omega_{\rm sup}$ of the system
that would provide rotational support. Characterizing the system by
its angular momentum $J \simeq \omega M R^2$ and rotational support by
$\omega^2_{\rm sup} R^2 \simeq GM/R$ one finds
\begin{equation}
\lambda =\frac{\omega}{\omega_{\rm sup}} = \frac{J}{G^{1/2}M^{3/2}R^{1/2}},
\label{lambda_paddy}
\end{equation}
which is up to a factor of $\sqrt{2}$ Bullock's $\lambda^\prime$
\citep{bullock:2001} if one replaces $GM/R$ by the circular velocity
$V^2_{\rm circ}$. Contrary to Eq.(\ref{lambda_peebles}) both Bullock's definition and Eq.(\ref{lambda_paddy})
contain only quantities which can be easily calculated in numerical
simulations for spherical halos. If we assume that the total energy of
the system is characterized by $E \simeq -GM^2/R$ one comes back to
Peebles' original definition.

The halos in the numerical simulation are assumed to be virialised and
thus characterized by $2T+U = 0$, where $T$ and $U$ are the kinetic
and potential energies of the object. \cite{bett:2006} show in their
Fig.  5 how real halos scatter around this assumption. Replacing the
potential energy in Eq.(\ref{lambda_peebles}) by $(-2T)$ we end up with a third
definition of the spin parameter:
\begin{equation}
\lambda =\frac{J T^{1/2}}{GM^{5/2}},
\label{lambdaT}
\end{equation}
As with the previous definition this $\lambda$ can be easily
calculated numerically. The advantage of this spin parameter is that
it can be calculated easily for any shape of the halo. Therefore, this
definition is especially suited for the calculation of the spin
parameter of halos found with the FOF analysis.

\begin{figure}
\epsscale{1.0}
\plotone{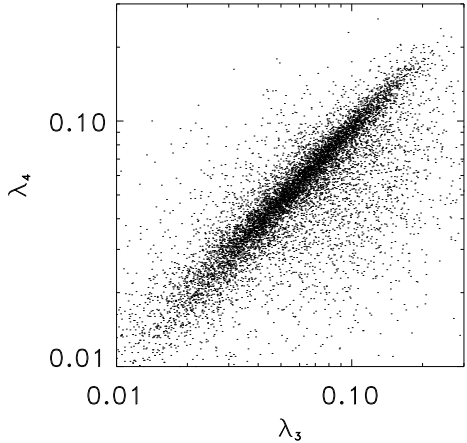}  
\caption{\label{plot:spin_spin}
  Comparison of the DM spin parameter $\lambda_3$ calculated according
  to Eq.  (\ref{lambdaT}) and $\lambda_4$  computed from  Eq.
  (\ref{lambda})}
\end{figure}

The halos identified in the MareNostrum universe consists of two
components, dark matter and gas.  Then for each component the spin
parameter is the ratio of the angular velocity $\omega$ of that
component to $\omega_{\rm sup}$ of the system. Therefore, we have
\begin{equation}
\lambda_{\rm gas(DM)} = \frac{J_{\rm gas(DM)}}
          {M_{\rm gas(DM)}(2 G (M_{\rm gas}+M_{\rm DM}) R_{\rm vir})^{1/2}},
\label{lambda}
\end{equation}
where $M_{\rm gas(DM)}$ denote the gas (DM) mass inside the virial
sphere of radius $R_{\rm vir}$ and $J_{\rm gas(DM)}$ the angular
momentum of the corresponding component. Note, that we have introduced
in Eq.(\ref{lambda}) an additional quotient of $\sqrt{2}$ to be
consistent with the definition of \cite{bullock:2001}.

\begin{figure*}
\plottwo{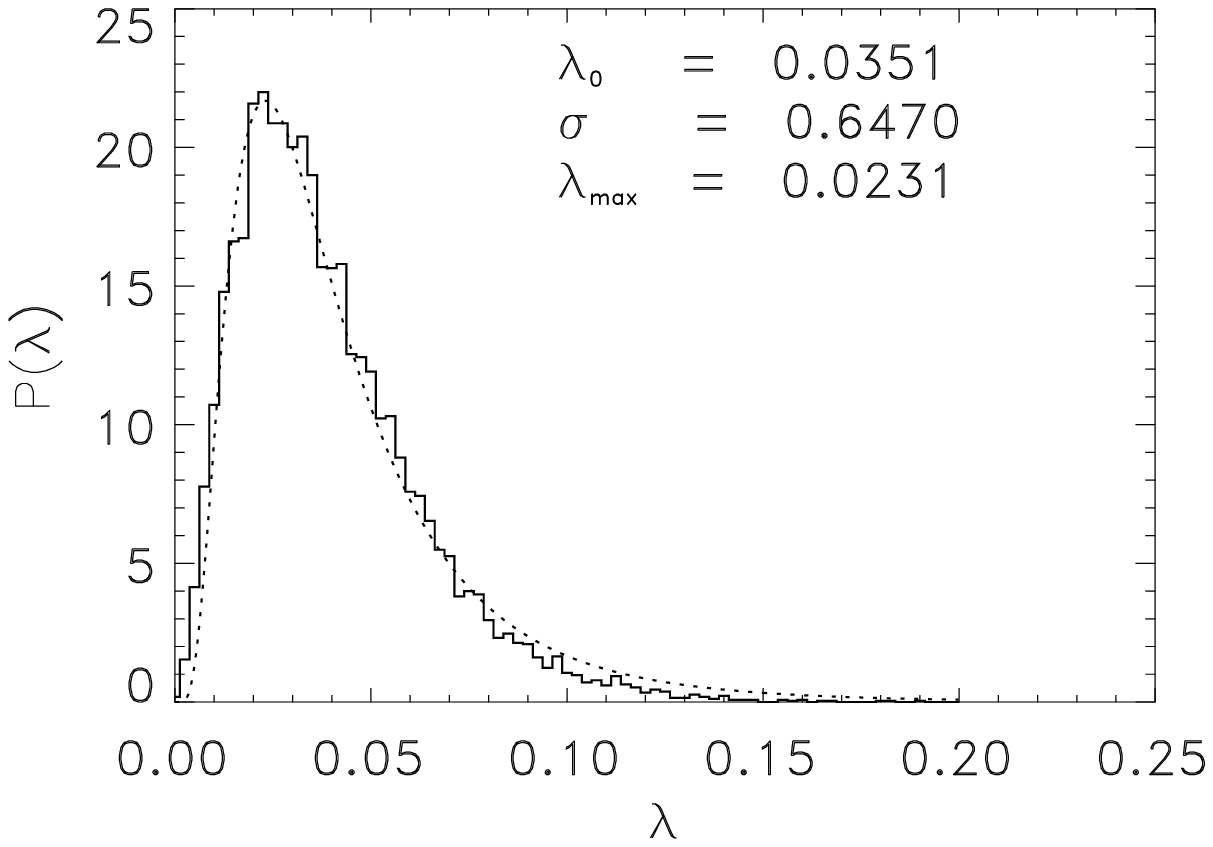}{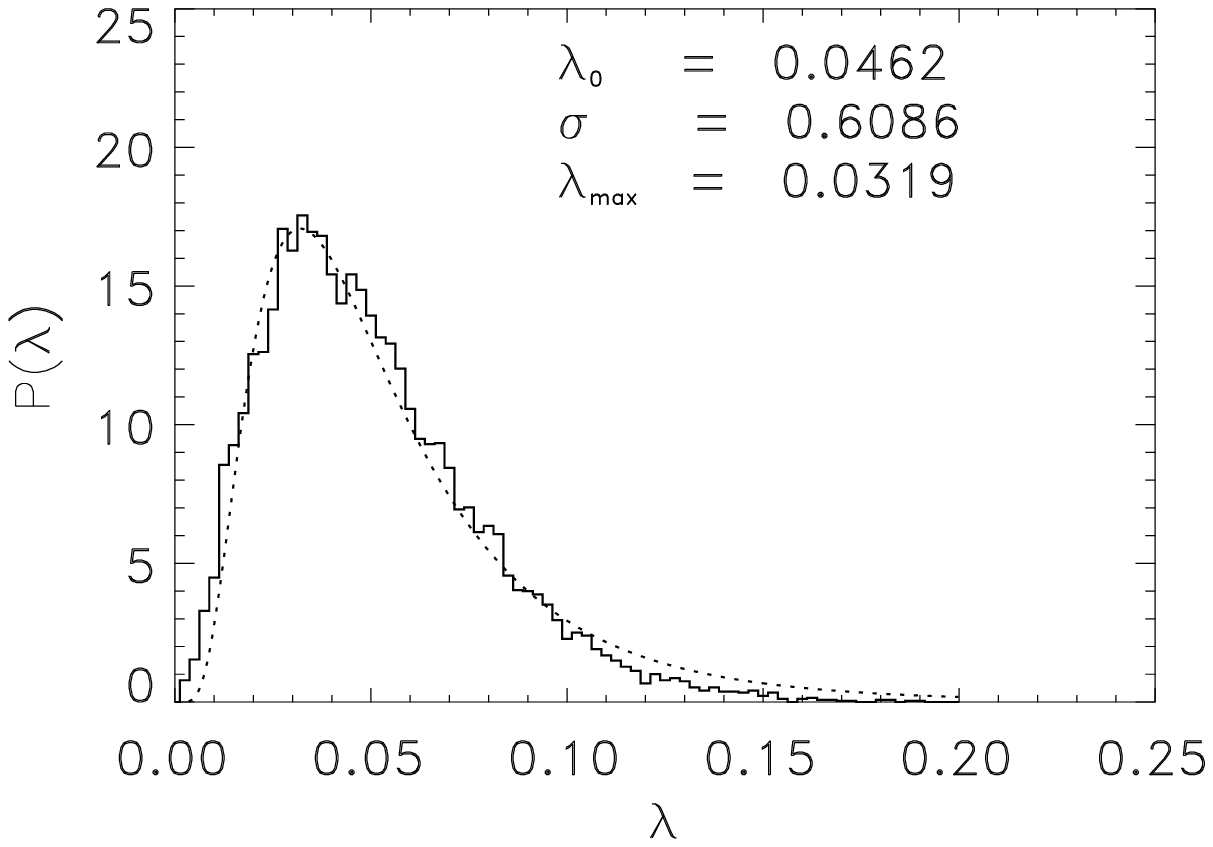}  
\caption{\label{plot:spin}
  {\it Left}: Distribution of the spin parameter of the dark matter in
  halos with masses larger than $5 \times 10^{13} \hMsun$.  The dotted
  line is a log-normal distribution Eq.(\ref{lognormal}) with the
  parameters $\lambda_0$ and $\sigma$ as given in the plot.  {\it
    Right}: The same for the spin parameter of the gas in halos.  }
\end{figure*}

\cite{vdbosch:2002} and \cite{sharma:2005} discussed in detail the
angular momentum distributions of the gas and dark matter components
of halos found in SPH cosmological simulations.  To this end they
decomposed the particle velocities into a streaming and a random
component. The velocities of the SPH gas particles are streaming
velocities to which one could add microscopic random motions.  The
velocities of the DM particles contain random motions which could be
eliminated by smoothing over a certain interpolation kernel.
Nevertheless, as long as the number of particles in the halo is very
large compared to the number of smoothing neighbors the effects of the
smoothing on the spin parameter can be neglected (Sharma: private
communication).  Following \cite{sharma:2005} we use the total
velocities of the DM and gas particles as provided by the simulation
to calculate the spin parameters of both components.

We want to compare the spin parameters of a given structure calculated
by the different abovementioned definitions.  The objects found by the
FOF algorithm have arbitrary shapes and the spin parameter as defined
in Eq.  (\ref{lambdaT}), $\lambda_3$ is calculated with respect to the
center of mass of each object. On the other hand, we have also
calculated the spin parameter for spherical halos at virial
overdensity ($334\; \times$ mean density at redshift $z=0$). In this
case, we first identify for each of the FOF objects the highest
density peak.  To this end, we decrease the linking length by a factor
of 8 ending up with sub-structures of about $170.000 \times$ mean
density.  For each FOF object we take the center of mass of the most
massive sub-structure as the center of a sphere and determine the
virial radius and mass within this sphere.  Then we use Eq.
(\ref{lambda}) to calculate the DM spin parameter $\lambda_4$.  In
Fig. \ref{plot:spin_spin} we compare the DM spin parameters
$\lambda_3$ and $\lambda_4$. Having in mind the completely different
treatments of the spin parameter they agree surprisingly well.  More
than 63 \% of our halos are within 20 \% scatter around the
$\lambda_3=\lambda_4$ relation. 
 The reason for the larger deviations shown in this plot is due to 
  the fact that halos defined from the spherical overdensity criterium and
  those defined from FOF algorithm look completely different if
  substructures are present. In such cases, the spin of the FOF
  objects is calculated with respect to the center of mass whereas for
  the spherical halos it is calculated with respect to the the
  position of the highest density peak. These positions, and the
  corresponding spins may differ substantially. For those halos in
  which this effect is not important, the spin parameters defined by
  both methods have very similar values. However, there is also
  another source of scatter, although much smaller than the previous
  one, due to the assumption of virial equilibrium $2T+U = 0$ which is
  not exactly fulfilled in our halos. Thus, even for spherical objects
  the spins $\lambda_3$ and $\lambda_4$ will not be identical.
In fact, \cite{shapiro:2004} pointed out that the presence of
infalling matter acts as a surface pressure even for collisionless DM.
The surface term leads to $2T+U > 0$. The spin parameter calculated
from the kinetic energy, $\lambda_3$ tends to be slightly larger than
the spin parameter $\lambda_4$, probably because the total kinetic
energy of DM particles is larger than that assumed by virial
equilibrium.  \cite{hetznecker:2006} have studied the distribution of
the virial coefficient $\eta=2T/|U|$ for a sample of dark matter halos
from $\Lambda$CDM N-body simulations and showed that the mean value of
$\eta$ approaches unity at redshift $z=0$.

In \S{} \ref {sec:shapes} we found different shapes (and therefore
volumes) for the DM and gas components of the halos.  Therefore, to
compare the spin parameter of both components we better used $\lambda$
defined by Eq. \ref{lambda}.  We put a sphere at the position of the
most massive substructure as described above and determine the virial
radius and mass of the halo as well as the gas and dark matter masses
(see \S{} \ref{sec:baryonfraction}) and the corresponding angular
momenta.

The resulting distribution of the spin parameter of the gas and dark
matter components are shown in Fig. \ref{plot:spin}. The distribution
of the spin parameter $\lambda$ can be described by a log-normal
distribution
\begin{equation}
P(\lambda)  = \frac{1}{\sqrt{2\pi}\sigma_{\lambda} \lambda} 
              \exp\left[ 
               -\frac{\ln^2(\lambda/\lambda_0)}{2\sigma^2_{\lambda}}
              \right] .
\label{lognormal}
\end{equation}
The best fit parameters are for the DM distribution $\lambda_0 =
0.0351 \pm 0.0016$, $\sigma_{\lambda} = 0.6470 \pm 0.0067$ and for
the gas distribution $\lambda_0 = 0.0462 \pm 0.0012$,
$\sigma_{\lambda} = 0.6086 \pm 0.0030$. $P(\lambda)$ has a maximum at
$\lambda_{max}= 0.0231$, 0.0319 for the DM resp. gas distribution.
Recently, \cite{bett:2006} have proposed another distribution function
which fits their data better. Since the scatter of $\lambda_0$ with
varying total number of halos depending on the lower mass cut-off is in
our case of the same order as the error in the determination of
$\lambda_0$ we believe that the log-normal function is a sufficient
fit to describe the behavior of the spin parameter. The differences
of $\lambda_0$ using different definitions of $\lambda$ are at least
of the same order.

Since dark matter dominates the total mass the total distribution of
the spin parameter of dark matter and gas practically coincides with
that of the dark matter only. Moreover, it is in agreement with the
distribution of the spin parameter calculated by
\cite{gottloeber:2006a} using Eq.(\ref{lambdaT}) for the non-spherical
FOF halos.

\begin{figure}
\plotone{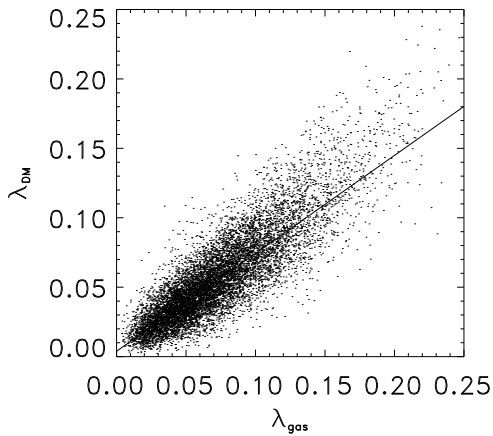}  
\caption{\label{plot:spin_gasDM}
  Distribution of the spin parameter of the gas and DM components. The
  solid line is the Least Square linear fit $\lambda_{DM}=0.72 \lambda_{gas} +
  0.004$.  }
\end{figure}

\cite{vdbosch:2002} found agreement between the spin distribution of
the gas and DM components of 378 halos identified in a small box of
$10 \hMpc$ at redshift $z=3$. We see a substantial shift of the
log-normal distribution of the gas spin towards higher spin in
comparison to the dark matter. Our $\lambda_0$ are slightly larger
than the values (0.0287, 0.0412) reported by \cite{sharma:2005} for a
sample of 41 halos in a box of $32.5 \hMpc$ at redshift $z=0$. We find
a mean $\lambda_{gas}/\lambda_{DM}$ of 1.39 (see Fig.
\ref{plot:spin_gasDM})
 with a tendency to decrease with halo mass. The
standard deviation is 0.57. Only the 159 most massive halos ($M_{vir}
> 5 \times 10^{14} \hMsun$) show a substantially smaller
$\lambda_{gas}/\lambda_{DM} = 1.23$ with a standard deviation of 0.45.

At $z=1$ we found a similar behavior of the spins of the DM and gas
components. Both spins follow log-normal distributions which can be
described by $\lambda_0 = 0.0467$ and 0.0541 respectively, {\em i.e.}
the mean spin of the gas component is 1.16 times larger than the spin
of the DM component. During further evolution this ratio increases
until 1.39. The spins themselves are slightly larger at redshift
$z=1$.

%%
%%------------------------------
\section{Baryon fraction}
\label{sec:baryonfraction}
%%------------------------------
%%

The observed cluster baryon fraction is an important tool for the
determination of cosmological parameters. Typically the gas fraction
in clusters is measured by X-ray observations at overdensities 500 or
larger, i.e. well inside the virial radius. A certain fraction of the
baryons reside in stars. Using a non-radiative simulation to determine
the baryon fraction is by sure a simplification, but since we are
interested in the baryon fraction at virial radius we expect not to be
affected very much by neglecting cooling and star formation processes
(see the discussion below).

During the last decade baryon fractions in clusters $Y_b = f_{\rm
  cluster}/f_{\rm cosmic} = f_{\rm cluster}/(\Omega_b/\Omega_{\rm
  matter})$ have been studied in non-radiative simulation by several
authors. \cite{eke:1998} found within the virial radius a baryon
fraction of 0.85 -- 0.9. Within the Santa Barbara cluster comparison
project \citep{frenk:1999} the baryon fraction at virial radius
averaged over all codes was $Y_b = 0.92$. They found a substantial
scatter between codes and a systematic offset between SPH and grid
codes which led to higher baryon fractions.  Recently,
\cite{kravtsov:2005} found that at large radii the baryon fraction in
the ART simulations is by about 3\% -- 5\% higher than in GADGET
simulations of the same clusters. Within our non-radiative simulation
we have explored the baryon fraction at the virial radius in objects
with virial masses larger than $5 \times 10^{13} \hMsun$. The assumed
virial overdensity at redshift $z=0$ is 334, at redshift $z=1$ it is
201. At redshifts $z = 0$ we found more than 10000 objects and about
2500 at redshift $z = 1$.  The scatter in the measured baryon fraction
of our cluster was found to be quite large.  At redshift $z=0$ it
ranges between 0.85 and 1.0 with a mean of 0.92
\citep{gottloeber:2006a}.

\begin{figure}
\plotone{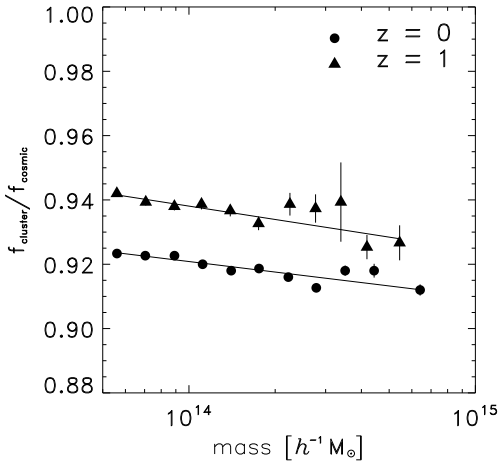}  
\caption{\label{plot:baryonfraction}
  Baryon fraction, normalizwed to the cosmic baryon fraction,  
 in clusters as a function of halo mass for two
  different redshifts.
}
\end{figure}

In Fig. {\ref {plot:baryonfraction}} we show the ratio $Y_b$ as a
function of the virial mass of clusters. Due to the large number of
objects the standard deviation of the mean value of $Y_b$ for
different mass bins is small and we could fit a linear relation in the
semi-logarithmic plot, $ Y_b = \alpha \ln(M) + \beta$, with the slope
$\alpha = -0.005 \pm 0.001$ at redshift $z=0$ and $\alpha = -0.006 \pm
0.001$ at redshift $z=1$. There is a 2\% decrease of the baryon
fraction between redshifts one and zero.  \cite{ettori:2006} found 5
\% decrease of the baryon fraction in this time interval.  Within our
model the decrease can be naturally expected. Since the gas works
against its pressure during the formation of the cluster the total
dark matter mass grows slightly faster than the gas mass. Therefore,
the relative gas fraction decreases with time as can be seen in Fig.
{\ref {plot:baryonfraction}}. For illustration we have selected 1200
clusters with masses larger than $2 \times 10^{14} \hMsun$ and 2900
clusters with masses smaller than $2 \times 10^{14} \hMsun$ and
compared the mass growth of their DM and baryonic components between
$z=0$ and $z=1$, $D_{DM,gas} = m_{DM,gas}(z=0)/ m_{DM,gas}(z=1)$. The
averaged relative growth $D_{DM} / D_{gas}$ is 1.03 for $M > 2 \times
10^{14} \hMsun$ and 1.02 for $M < 2 \times 10^{14} \hMsun$, i.e.
clusters accrete DM faster than gas and the mean baryonic fraction
decreases with time. Originally the baryon fraction was homogenous.
During collapse of the whole cluster region the baryonic fraction
decreases slowly. Most massive clusters are situated in the positions
of the highest density peaks in the originally almost homogenous
medium. These peaks start to grow first. This could be the reason for
small but statistically significant slope $\alpha$ which we found.
Note, however, that recently \cite{crain:2006} have found the
opposite, a slightly increasing baryon fraction with mass, from a set
of clusters extracted from a non-radiative SPH re-simulation of the
Millenium Run, but with 2 times less number of particles than ours.
Also, the relative baryon fraction of their clusters (at overdensity
200) falls below 0.9.

Recently \cite{kravtsov:2005} studied the effects of cooling and star
formation on the the baryon fractions in clusters. They showed that
the cooling of the gas and the associated star formation increases the
baryon fraction even at radii as large as the virial radius. The
averaged baryon fraction at virial radius differ by 5 \% between the
simulations with cooling and star formation ($Y_b=1.02$)  and the
non-radiative simulations ($Y_b=0.97$), which is already $\sim  5$ \%
larger than the value we obtained from the MareNostrum Universe SPH
simulation. \cite{ettori:2006} have also  studied a set of 
clusters obtained both in non-radiative SPH simulations as well as in
simulations  including cooling and star formation. 
For the non-radiative simulations they found an averaged
baryon fraction  of 0.89, which is  slightly lower than the  value we
found. In agreement with \cite{kravtsov:2005} they also  found in their
simulations with cooling and star formation that the baryon fraction
 at virial radius increases  with respect to the non-radiative case by
 about  3 \%.

%%
%%------------------------------
\section{CONCLUSIONS}
\label{sec:sum}
%%------------------------------
%%

We have selected a statistically significant sample of more than 
10,000 clusters from the MareNostrum universe simulation. With this
database we have determined  the shape and the spin of the DM and gas
components of the clusters independently. We found that both the gas
and dark matter components tend to be prolate although the gas is much
more spherically distributed.  The mean ratio of the minor to major
axis decreases with increasing halo mass, i.e. younger objects tend to
be less spherical, in agreement with the  results of \cite{kasun:2005}
obtained from the Hubble Volume DM only simulation. 
On the other hand,  
the triaxiality parameters of the gas and DM component 
differ  by a few percent only and both of them increase with
halo mass. 

The spin parameters of the DM and gas components are well represented
by lognormal distribution functions with $\lambda_0 = 0.0351$ (DM) and
$\lambda_0 = 0.0462$ (gas).  These values are slightly larger than the
values found for a sample of 41 halos with masses from dwarf to bright
galaxies \citep{sharma:2005}. On average, the spin of the gas
component is by a factor of 1.39 larger than the spin of the DM
component.  This ratio is smaller for very massive halos, 1.23 for
halos with $M_{vir} > 5 \times 10^{14} \hMsun$.  It decreases slightly
with redshift being 1.16 at $z=1$.

On average, the baryon fraction in the cluster sample is $Y_b = 0.92$.
The baryon fraction increases with redshift, being 0.94 at $z=1$. At
both redshift the baryon fraction decreases slightly with increasing
virial mass of the clusters. 
The radiative processes of cooling and star formation are
  expected to change the baryon content found for halos in
  non-radiative simulations. There is a discrepancy between the
  predicted baryon fraction in SPH and AMR codes. Both effects are of
  the same order as the time and mass dependence found here. Thus, at
  present observational projects to estimate cosmological parameter
  such as the total matter density in the universe and the equation of
  state of the dark energy can still safely assume a rather universal,
  unevolving baryon fraction for clusters of galaxies.

\section*{Acknowledgments} 

The {\sc MareNostrum Universe} has been created at the Barcelona
Supercomputer Center and analyzed at NIC J\"ulich.  We thank Acciones
Integradas Hispano-Alemanas and DFG for financial support. GY would
like to thank also MCyT for financial support under project numbers
AYA2003-07468 and BFM2003-01266. We appreciate very much the fruitful
discussions with S. Sharma, M. Hoeft, M. Steinmetz and A.  Knebe.  We
would also like to thank the anonymous referee for her/his valuable
comments on the first version of the paper.

\end{document}